\documentclass[floatfix,aps,superscriptaddress,prd,amsmath,nofootinbib,preprintnumbers,twocolumn]{revtex4}

\usepackage{graphicx}
\usepackage{bm}
\usepackage[
colorlinks=true,        
citecolor=blue,         
linkcolor=blue,         
urlcolor=blue           
]{hyperref}             

\newcommand{\nc}{\newcommand*}
\nc{\Om}{\Omega}
\nc{\ogw}{\Omega_{\mathrm{GW}}}
\nc{\rd}{\mathrm{d}}
\nc{\eg}{\textit{e.g.~}}
\nc{\red}[1]{\textcolor{red}{#1}}
\nc{\lvc}{LIGO/Virgo} 

\def\({\left(}
\def\){\right)}
\def\[{\left[}
\def\]{\right]}

\def\e{\begin{equation}}
\def\q{\end{equation}}
\def\m{\begin{eqnarray}}
\def\n{\end{eqnarray}}

\begin{document}

\title{Measuring the scalar induced gravitational waves from observations}

\author{Jun Li}
\email{lijun@qust.edu.cn}
\affiliation{School of Mathematics and Physics,
    Qingdao University of Science and Technology,
    Qingdao 266061, China}
\affiliation{CAS Key Laboratory of Theoretical Physics,
    Institute of Theoretical Physics, Chinese Academy of Sciences,
    Beijing 100190, China}

\author{Guang-Hai Guo}
\email{ghguo@qust.edu.cn}
\affiliation{School of Mathematics and Physics,
    Qingdao University of Science and Technology,
    Qingdao 266061, China}

\date{\today}

\begin{abstract}

We measure the scalar induced gravitational waves from the cosmic microwave background (CMB) observations and the gravitational wave observations. In the $\Lambda$CDM+$r$ model, we constrain the cosmological parameters within the evolution of the scalar induced gravitational waves by the additional scalar power spectrum. The two special cases called narrow power spectrum and wide power spectrum have influence on the cosmological parameters, especially the combinations of Planck18+BAO+BK15+LISA. We also compare these numerical results from four datasets within LIGO, LISA, IPTA and FAST projects, respectively. The constraints from FAST have a significant impact on tensor-to-scalar ratio.
\end{abstract}

\maketitle


\section{Introduction}
Modern cosmology is based on general relativity and the premise that the Universe is homogeneous isotropic on large scale. On this background, the early Universe was highly inhomogeneous on small scale and the perturbations would evolve into stars, galaxies and galaxy clusters over time. The cosmological linear perturbation theory has been developed so fast these decades owning to the cosmological observations, such as the cosmic microwave background (CMB). The theoretical first-order approximation from perturbations has been revealed through the CMB observations which include the Wilkinson Microwave Anisotropy Probe (WMAP) \cite{Komatsu:2008hk}, the Planck satellite \cite{Aghanim:2018eyx}, the BICEP2 and Keck array (BK) \cite{Ade:2018gkx}, and the Baryon Acoustic Oscillation (BAO) \cite{Beutler:2011hx,Ross:2014qpa,Alam:2016hwk}. Depend on the cosmological linear perturbation theory and CMB observations, the standard cosmological model has been established and corresponding cosmological parameters are measured accurately.

However, the rapid developments of the cosmological observations inspire us considering the deviation from the first-order approximation. The second-order cosmological perturbation theory has been proposed in literatures \cite{Kenji:1967}. The result shows that the curvature perturbations couple to the tensor perturbations at second-order which produce the scalar induced gravitational waves in the radiation dominated era. The power spectrum from the curvature perturbations influences the evolution of the scalar induced gravitational waves. We should consider the scalar induced gravitational waves from theoretical  and observational sides. Applications of the second-order cosmological perturbation theory to the analysis of the cosmological observations become significant. See some other related works in \cite{Nakamura:2019zbe,Yuan:2019wwo,Yuan:2019fwv,Matarrese:1997ay,Kohri:2018awv,Ananda:2006af}.

In this paper, we add the scalar power spectrum as presented in a previous study \cite{Yuan:2019wwo} and measure the scalar induced gravitational waves from the CMB observations and the gravitational wave observations. Here, the observations include the Planck satellite, the BICEP2 and Keck array, the Baryon Acoustic Oscillation, the LIGO detector \cite{TheLIGOScientific:2016dpb, LIGOScientific:2019vic, Thrane:2013oya}, the LISA detector \cite{Thrane:2013oya, Caprini:2015zlo} and two Pulsar timing array projects \cite{Hellings:1983fr, Verbiest:2016vem, Nan:2011um, Kuroda:2015owv}, namely IPTA \cite{Verbiest:2016vem} and FAST \cite{Nan:2011um}. The gravitational wave observations are taken as our previous works \cite{Li:2019vlb,1842066}.

\section{The power spectrum and the scalar induced gravitational waves}
The second-order perturbation metric about the Freedmann-Robert-Walker background in the conformal Newtonian gauge is taken as \cite{Kohri:2018awv,Ananda:2006af}
\e
ds^2=a^2\left\{-(1+2\phi)d\eta^2+\left[(1-2\phi)\delta_{ij}+\frac{h_{ij}}{2}\right]dx^idx^j \right\},      \label{metric}
\q
where $\eta$ is the conformal time, $\phi$ is the scalar perturbation and $h_{ij}$ is the gravitational wave perturbation.

In the radiation dominated Universe, the fraction of the gravitational wave energy density per logarithmic wavelength is given by \cite{Kohri:2018awv}
\e
\Omega_{\mathrm{GW}}(\eta, k)=\frac{1}{24}\left( \frac{k}{a(\eta)H(\eta)}\right)^2\overline{P_h(\eta, k)},
\q
where $P_h(\eta, k)$ is the power spectrum of the induced gravitational waves
\e
P_h(\eta, k)=4\int_0^\infty\mathrm{d}v\int_{\vert1-v\vert}^{1+v}\mathrm{d}uf^2(v, u, x)P_{\zeta}(kv)P_{\zeta}(ku),
\q
and $P_{\zeta}(k)$ is the power spectrum of the primordial curvature perturbations. The function $f(v, u, x)$ is defined as
\e
f(v, u, x)=I(v, u, x)\frac{4v^2-(1+v^2-u^2)^2}{4vu},
\q
where function $I(v, u, x)$ comes from the source term.

Here, we consider adding the scalar power spectrum which is peaked at $k_*$ and parameterized as \cite{Yuan:2019wwo} 
\e
P_{\zeta}(k)=\left\{
\begin{aligned}
&\frac{k-k_{-}}{k_*-k_{-}}, \quad \mathrm{for} \ k_{-}<k<k_{*},    \\
&\frac{k_{+}-k}{k_{+}-k_{*}}, \quad \mathrm{for} \ k_{*}<k<k_{+},   
\end{aligned}
\right.
\q
where choose $k_*=10^{15} \ \mathrm{Mpc}^{-1}$ with the same reason as our previous work \cite{Li:2018iwg}. The amplitude of the additional scalar power spectrum is much larger than those at CMB scales which provide a new way to probe primordial black holes. We also introduce the parameter $\Delta$ to quantify the width of the additional scalar power spectrum
\e
\Delta=\frac{k_{+}-k_{-}}{k_*},
\q
and discuss two simple examples. The narrow power spectrum with $k_{-}=0.995k_*$ and $k_{+}=1.005k_*$ takes the form
\e
P_{\zeta}(k)=\left\{
\begin{aligned}
&\frac{k-0.995*10^{15}}{0.005*10^{15}}, \ \mathrm{for} \ 0.995*10^{15}<k<10^{15},    \\
&\frac{1.005*10^{15}-k}{0.005*10^{15}}, \ \mathrm{for} \ 10^{15}<k<1.005*10^{15}. \label{narrow}
\end{aligned}
\right.
\q
The wide power spectrum with $k_{-}=0.5k_*$ and $k_{+}=10.5k_*$ takes the form
\e
P_{\zeta}(k)=\left\{
\begin{aligned}
&\frac{k-0.5*10^{15}}{0.5*10^{15}}, \ \mathrm{for} \ 0.5*10^{15}<k<10^{15},    \\
&\frac{10.5*10^{15}-k}{9.5*10^{15}}, \ \mathrm{for} \ 10^{15}<k<10.5*10^{15}.  \label{wide}
\end{aligned}
\right.
\q
Then we will measure the scalar induced gravitational waves from the CMB observations and the gravitational wave observations as below.

\section{Constraints on the cosmological parameters}
\begin{figure*}
\centering
\includegraphics[width=16.2cm]{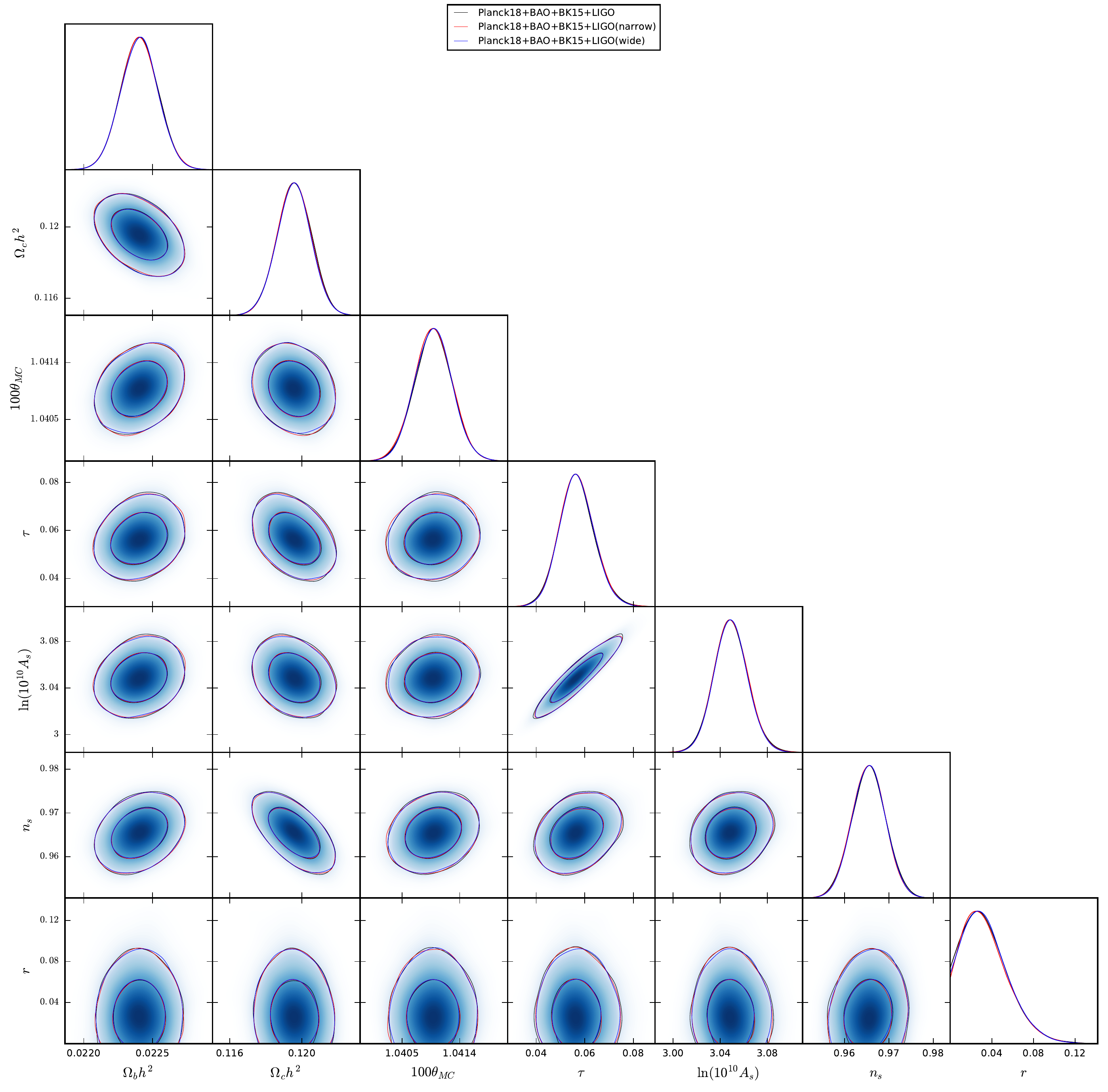}
\caption{The contour plots and the likelihood distributions for the cosmological parameters in the $\Lambda$CDM+$r$ model at the $68\%$ and $95\%$ CL from the combinations of Planck18+BAO+BK15+LIGO, Planck18+BAO+BK15+LIGO$^n$, Planck18+BAO+BK15+LIGO$^w$, respectively.}
\label{fig:picture1}
\end{figure*}

\begin{figure*}
\centering
\includegraphics[width=16.2cm]{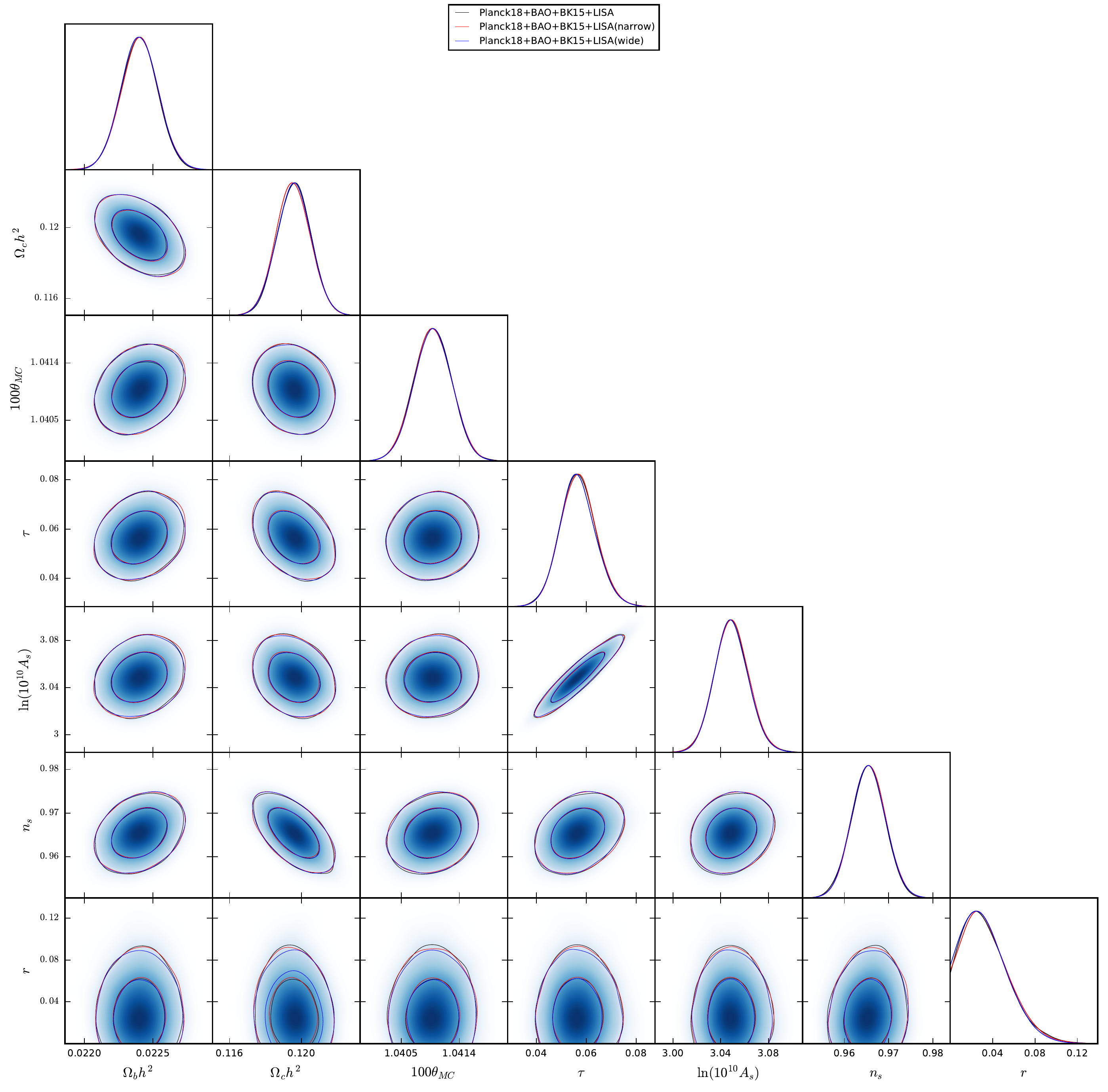}
\caption{The contour plots and the likelihood distributions for the cosmological parameters in the $\Lambda$CDM+$r$ model at the $68\%$ and $95\%$ CL from the combinations of Planck18+BAO+BK15+LISA, Planck18+BAO+BK15+LISA$^n$, Planck18+BAO+BK15+LISA$^w$, respectively.}
\label{fig:picture2}
\end{figure*}

\begin{figure*}
\centering
\includegraphics[width=16.2cm]{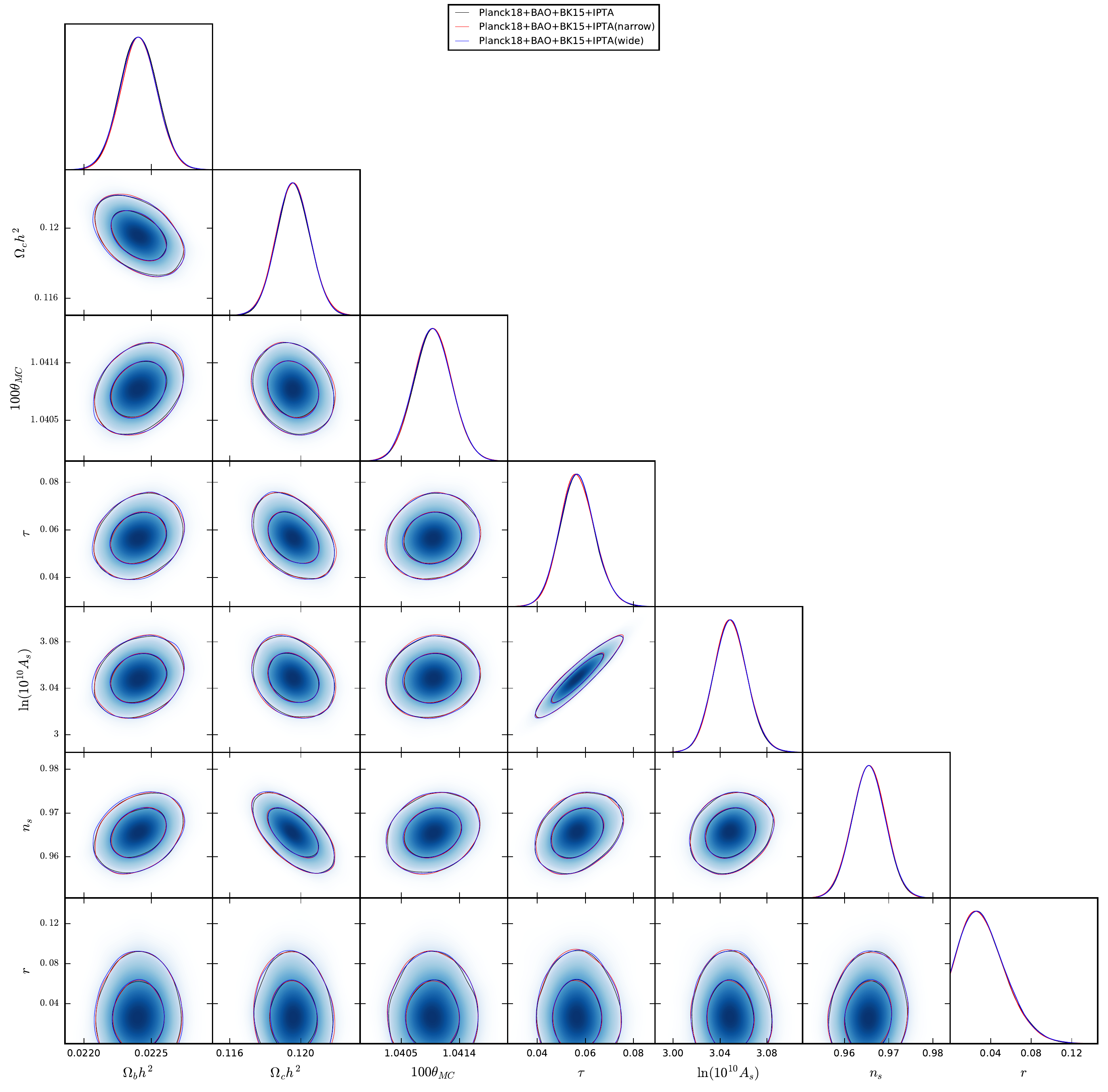}
\caption{The contour plots and the likelihood distributions for the cosmological parameters in the $\Lambda$CDM+$r$ model at the $68\%$ and $95\%$ CL from the combinations of Planck18+BAO+BK15+IPTA, Planck18+BAO+BK15+IPTA$^n$, Planck18+BAO+BK15+IPTA$^w$, respectively.}
\label{fig:picture3}
\end{figure*}

\begin{figure*}
\centering
\includegraphics[width=16.2cm]{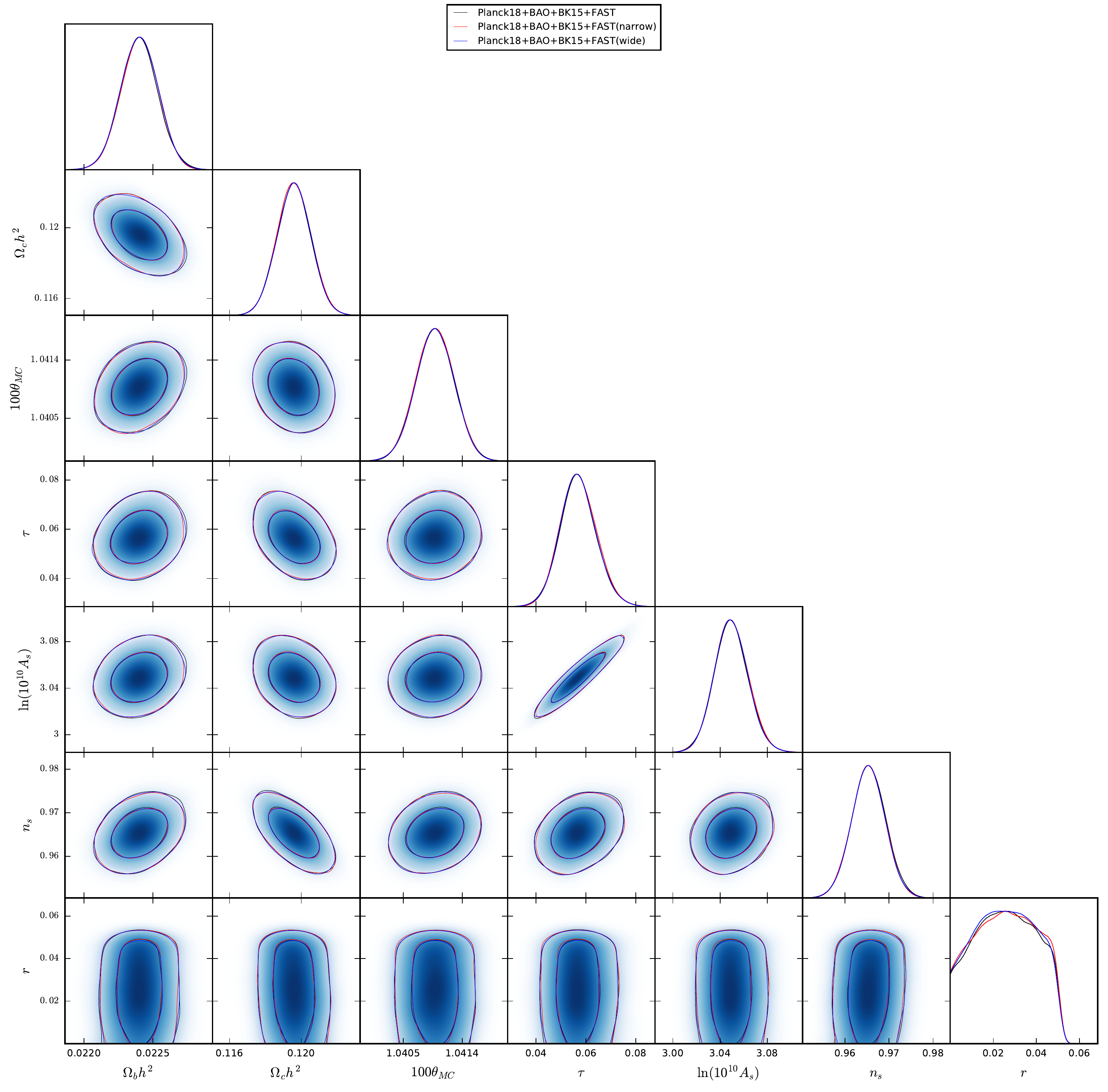}
\caption{The contour plots and the likelihood distributions for the cosmological parameters in the $\Lambda$CDM+$r$ model at the $68\%$ and $95\%$ CL from the combinations of Planck18+BAO+BK15+FAST, Planck18+BAO+BK15+FAST$^n$, Planck18+BAO+BK15+FAST$^w$, respectively.}
\label{fig:picture4}
\end{figure*}

\begin{figure*}
\centering
\includegraphics[width=16.2cm]{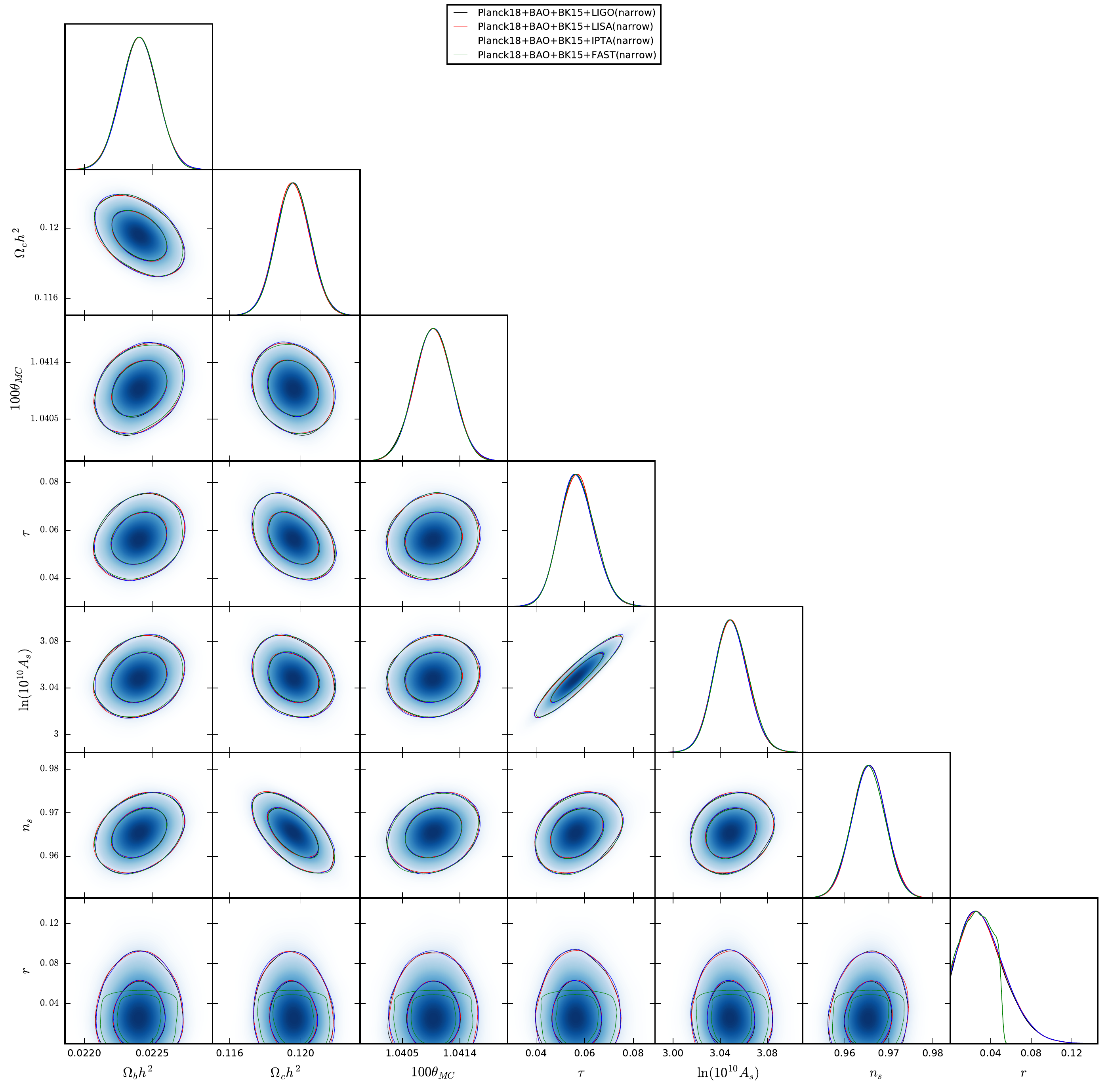}
\caption{The contour plots and the likelihood distributions for the cosmological parameters in the $\Lambda$CDM+$r$ model at the $68\%$ and $95\%$ CL from the combinations of Planck18+BAO+BK15+LIGO$^n$, Planck18+BAO+BK15+LISA$^n$, Planck18+BAO+BK15+IPTA$^n$, Planck18+BAO+BK15+FAST$^n$, respectively.}
\label{fig:picture5}
\end{figure*}

\begin{figure*}
\centering
\includegraphics[width=16.2cm]{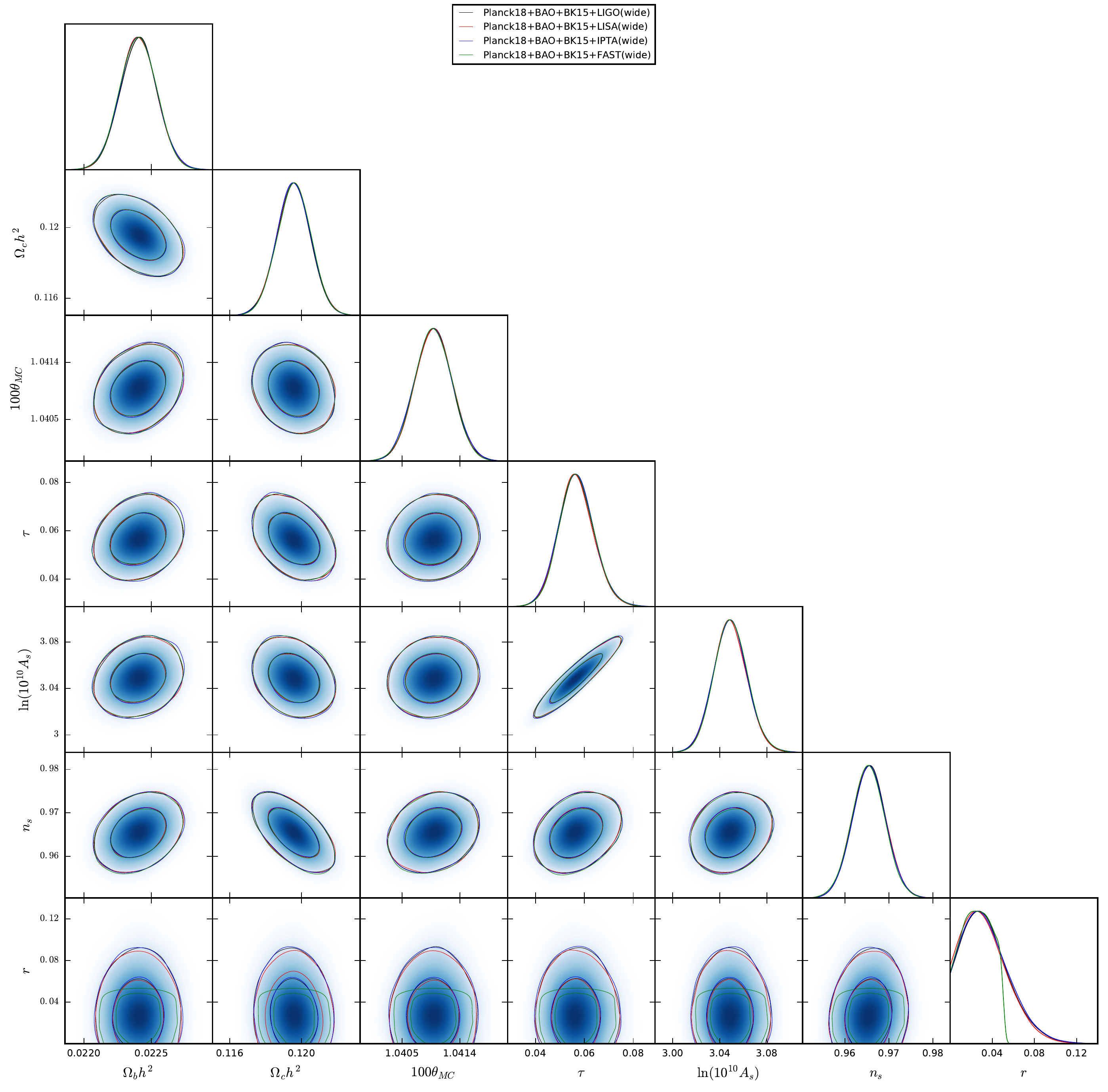}
\caption{The contour plots and the likelihood distributions for the cosmological parameters in the $\Lambda$CDM+$r$ model at the $68\%$ and $95\%$ CL from the combinations of Planck18+BAO+BK15+LIGO$^w$, Planck18+BAO+BK15+LISA$^w$, Planck18+BAO+BK15+IPTA$^w$, Planck18+BAO+BK15+FAST$^w$, respectively.}
\label{fig:picture6}
\end{figure*}

We use the publicly available codes Cosmomc \cite{Lewis:2002ah} to constrain the cosmological parameters within the evolution of the scalar induced gravitational waves by the additional scalar power spectrum.
In the standard $\Lambda$CDM model, the six parameters are the baryon density parameter $\Omega_b h^2$, the cold dark matter density $\Omega_c h^2$, the angular size of the horizon at the last scattering surface $\theta_\text{MC}$, the optical depth $\tau$, the scalar amplitude $A_s$ and the scalar spectral index $n_s$. We extend this model by adding the parameter $r$ and consider these seven parameters as fully free parameters. The parameter $r$, called tensor-to-scalar ratio, quantify the tensor amplitude $A_t$ compared to the scalar amplitude $A_s$ at the pivot scale
\e
r\equiv\frac{A_t}{A_s}.
\q
The consistency relation is ignored for more inflation models. Our numerical results are given in Table.~\ref{table:table1} to Table.~\ref{table:table2} and Fig.~\ref{fig:picture1} to Fig.~\ref{fig:picture6}.

\begin{table*}
\newcommand{\tabincell}[2]{\begin{tabular}{@{}#1@{}}#2\end{tabular}}
  \centering
  \begin{tabular}{ c | c| c| c |c| c| c}
  \hline
  \hline
  Parameters & \tabincell{c}{Planck18+BAO\\+BK15+LIGO} & \tabincell{c}{Planck18+BAO\\+BK15+LIGO$^n$} & \tabincell{c}{Planck18+BAO\\+BK15+LIGO$^w$} & \tabincell{c}{Planck18+BAO\\+BK15+LISA} & \tabincell{c}{Planck18+BAO\\+BK15+LISA$^n$} & \tabincell{c}{Planck18+BAO\\+BK15+LISA$^w$} \\
  \hline
  $\Omega_bh^2$ &  $0.02240^{+0.00014}_{-0.00013}$ & $0.02240\pm0.00013$ & $0.02240\pm0.00013$ &  $0.02240\pm0.00013$ & $0.02240\pm0.00013$ & $0.02240\pm0.00013$\\
  $\Omega_ch^2$ &  $0.11958\pm0.00096$ & $0.11956\pm{0.00094}$ & $0.11956\pm0.00094$  &  $0.11959^{+0.00094}_{-0.00095}$ & $0.11954\pm{0.00093}$ & $0.11957^{+0.00094}_{-0.00095}$ \\
  $100\theta_{\mathrm{MC}}$ & $1.04099\pm0.00029$ & $1.04098^{+0.00030}_{-0.00029}$ & $1.04099\pm0.00029$ & $1.04099\pm0.00029$ & $1.04099\pm{0.00029}$ & $1.04099\pm0.00029$\\
  $\tau$ &   $0.0567^{+0.0069}_{-0.0078}$ & $0.0568^{+0.0070}_{-0.0078}$ & $0.0568^{+0.0069}_{-0.0076}$ &   $0.0567^{+0.0069}_{-0.0076}$ & $0.0569^{+0.0069}_{-0.0077}$ & $0.0567^{+0.0069}_{-0.0077}$\\
  $\ln\(10^{10}A_s\)$  & $3.049^{+0.014}_{-0.015}$ &$3.049\pm{0.014}$ & $3.049\pm0.014$  & $3.049^{+0.014}_{-0.015}$ &$3.049\pm{0.014}$ & $3.049\pm0.014$\\
  $n_s$ &  $0.9654\pm0.0039$  & $0.9654\pm{0.0038}$ & $0.9655\pm0.0038$ &  $0.9654^{+0.0037}_{-0.0038}$  & $0.9655\pm{0.0037}$ & $0.9655\pm0.0038$\\
  $r_{0.05}$ ($95\%$ CL) & $<0.076$ & $<0.075$ & $<0.074$  & $<0.075$ & $<0.075$ & $<0.073$ \\
  \hline
  \hline
  \end{tabular}
  \caption{The $68\%$ limits on the cosmological parameters in the $\Lambda$CDM+$r$ model from the combinations of Planck18+BAO+BK15+LIGO, Planck18+BAO+BK15+LIGO$^n$, Planck18+BAO+BK15+LIGO$^w$ and the combinations of Planck18+BAO+BK15+LISA, Planck18+BAO+BK15+LISA$^n$, Planck18+BAO+BK15+LISA$^w$, respectively.}
  \label{table:table1}
\end{table*}

In the $\Lambda$CDM+$r$ model, the combinations of LIGO and Planck18+BAO+BK15 constrain the seven parameters accurately which have presented in \cite{1842066}. When we consider the narrow power spectrum and the wide power spectrum as Eq.~(\ref{narrow}) and Eq.~(\ref{wide}), the corresponding numerical results of Planck18+BAO+BK15+LIGO$^n$ and Planck18+BAO+BK15+LIGO$^w$ change slightly. The effects from the scalar induced gravitational waves on the cosmological parameters are not obvious. 
The constraint on parameter $r$ is
\m
r &<& 0.076\quad(95\% \ \mathrm{C.L.}),
\n
from Planck18+BAO+BK15+LIGO datasets;
\m
r &<& 0.075\quad(95\% \ \mathrm{C.L.}),
\n
from Planck18+BAO+BK15+LIGO$^n$ datasets;
\m
r &<& 0.074\quad(95\% \ \mathrm{C.L.}),
\n
from Planck18+BAO+BK15+LIGO$^w$ datasets which are given in Table.~\ref{table:table1} and Fig.~\ref{fig:picture1}.

The combinations of Planck18+BAO+BK15+LISA$^n$ and Planck18+BAO+BK15+LISA$^w$ constrain the seven parameters better which is obvious from the contour plot of the cold dark matter density $\Omega_c h^2$ and the tensor-to-scalar ratio $r$ in Fig.~\ref{fig:picture2}.
The constraint on parameter $r$ is
\m
r &<& 0.075\quad(95\% \ \mathrm{C.L.}),
\n
from Planck18+BAO+BK15+LISA datasets;
\m
r &<& 0.075\quad(95\% \ \mathrm{C.L.}),
\n
from Planck18+BAO+BK15+LISA$^n$ datasets;
\m
r &<& 0.073\quad(95\% \ \mathrm{C.L.}),
\n
from Planck18+BAO+BK15+LISA$^w$ datasets.

\begin{table*}
\newcommand{\tabincell}[2]{\begin{tabular}{@{}#1@{}}#2\end{tabular}}
  \centering
  \begin{tabular}{ c | c| c| c| c| c| c}
  \hline
  \hline
  Parameters & \tabincell{c}{Planck18+BAO\\+BK15+IPTA} & \tabincell{c}{Planck18+BAO\\+BK15+IPTA$^n$} & \tabincell{c}{Planck18+BAO\\+BK15+IPTA$^w$} & \tabincell{c}{Planck18+BAO\\+BK15+FAST} & \tabincell{c}{Planck18+BAO\\
  +BK15+FAST$^n$} & \tabincell{c}{Planck18+BAO\\+BK15+FAST$^w$} \\
  \hline
  $\Omega_bh^2$ &  $0.02241\pm0.00013$ & $0.02240\pm0.00013$ & $0.02240\pm0.00013$ &  $0.02240\pm0.00013$ & $0.02240\pm0.00013$ & $0.02240\pm0.00014$ \\
  $\Omega_ch^2$ &  $0.11957\pm0.00093$ & $0.11957^{+0.00094}_{-0.00095}$ & $0.11957^{+0.00095}_{-0.00093}$ &  $0.11958\pm0.00094$ & $0.11959\pm0.00094$ & $0.11960\pm0.00094$   \\
  $100\theta_{\mathrm{MC}}$ & $1.04099\pm0.00029$ & $1.04099\pm{0.00029}$ & $1.04099\pm0.00030$ & $1.04099\pm0.00029$ & $1.04098\pm{0.00029}$ & $1.04099\pm0.00029$\\
  $\tau$ &   $0.0568^{+0.0069}_{-0.0077}$ & $0.0568^{+0.0069}_{-0.0078}$ & $0.0568^{+0.0071}_{-0.0078}$  &   $0.0569^{+0.0070}_{-0.0078}$ & $0.0570^{+0.0071}_{-0.0078}$ & $0.0569^{+0.0069}_{-0.0077}$\\
  $\ln\(10^{10}A_s\)$  & $3.049\pm0.014$ &$3.049^{+0.014}_{-0.015}$ & $3.049\pm0.014$  & $3.049^{+0.014}_{-0.015}$ &$3.050^{+0.014}_{-0.015}$ & $3.049\pm0.014$\\
  $n_s$ &  $0.9655\pm0.0037$  & $0.9654\pm{0.0038}$ & $0.9655^{+0.0037}_{-0.0038}$ &  $0.9654\pm0.0038$  & $0.9653\pm{0.0038}$ & $0.9653\pm0.0037$\\
  $r_{0.05}$ ($95\%$ CL) & $<0.075$ & $<0.075$ & $<0.075$ & $<0.049$ & $<0.049$ & $<0.049$\\
  \hline
  \hline
  \end{tabular}
  \caption{The $68\%$ limits on the cosmological parameters in the $\Lambda$CDM+$r$ model from the combinations of Planck18+BAO+BK15+IPTA, Planck18+BAO+BK15+IPTA$^n$, Planck18+BAO+BK15+IPTA$^w$ and the combinations of Planck18+BAO+BK15+FAST, Planck18+BAO+BK15+FAST$^n$, Planck18+BAO+BK15+FAST$^w$, respectively.}
  \label{table:table2}
\end{table*}

The combinations of Planck18+BAO+BK15+IPTA$^n$ and Planck18+BAO+BK15+IPTA$^w$ effect the seven parameters slightly which is given in Table.~\ref{table:table2} and Fig.~\ref{fig:picture3}.
The constraint on parameter $r$ is
\m
r &<& 0.075\quad(95\% \ \mathrm{C.L.}),
\n
from Planck18+BAO+BK15+IPTA datasets;
\m
r &<& 0.075\quad(95\% \ \mathrm{C.L.}),
\n
from Planck18+BAO+BK15+IPTA$^n$ datasets;
\m
r &<& 0.075\quad(95\% \ \mathrm{C.L.}),
\n
from Planck18+BAO+BK15+IPTA$^w$ datasets. The influence on the cosmological parameters from the additional scalar power spectrum are not obvious.

The combinations of Planck18+BAO+BK15+FAST$^n$ and Planck18+BAO+BK15+FAST$^w$ effect the seven parameters slightly. The likelihood distributions for the tensor-to-scalar ratio change in Fig.~\ref{fig:picture4}.
The constraint on parameter $r$ is
\m
r &<& 0.049\quad(95\% \ \mathrm{C.L.}),
\n
from Planck18+BAO+BK15+FAST datasets;
\m
r &<& 0.049\quad(95\% \ \mathrm{C.L.}),
\n
from Planck18+BAO+BK15+FAST$^n$ datasets;
\m
r &<& 0.049\quad(95\% \ \mathrm{C.L.}),
\n
from Planck18+BAO+BK15+FAST$^w$ datasets.

We also compare the narrow power spectrum and the wide power spectrum results in Fig.~\ref{fig:picture5} and Fig.~\ref{fig:picture6} from four datasets within LIGO, LISA, IPTA and FAST projects, respectively. The results show that the constraint from FAST have a significant impact on tensor-to-scalar ratio.

\section{summary}
In the $\Lambda$CDM+$r$ model, we constrain the scalar induced gravitational waves from the cosmic microwave background observations and the gravitational wave observations. The cosmological parameters are influenced by the evolution of the scalar induced gravitational waves from the additional scalar power spectrum, especially the combinations of Planck18+BAO+BK15+LISA. We also compare these numerical results from four datasets within LIGO, LISA, IPTA and FAST projects, respectively. The results show that FAST project has obvious impact on  tensor-to-scalar ratio.



\end{document}